\documentclass[letter]{aa} % for the letters 
\usepackage{graphicx}
\usepackage{mathtools}
\usepackage{txfonts}
\begin{document} 
\newcommand{\dg}{$^{\circ}$}
\newcommand{\bb}{2014 UN$_{271}$}

   \title{Size and albedo of the largest detected Oort-cloud object: comet C/2014 UN$_{271}$ (Bernardinelli-Bernstein)}

%   \subtitle{I. Overviewing the $\kappa$-mechanism}
\author{E. Lellouch\inst{1}
\and R.~Moreno\inst{1}
\and D. Bockel\'ee-Morvan\inst{1}
\and N.~Biver\inst{1}
\and P.~Santos-Sanz\inst{2}
}

 \institute{LESIA, Observatoire de Paris, PSL Research University, CNRS, Sorbonne Universit\'e, UPMC Univ. Paris 06, Univ. Paris Diderot, Sorbonne Paris Cit\'e, 5 place Jules Janssen, 92195 Meudon, France\\ %1
\email{emmanuel.lellouch@obspm.fr}
\and Instituto de Astrof\'isica de Andaluc\'ia-CSIC, Glorieta de la Astronom\'ia s/n, 18008-Granada, Spain.
            }

   \titlerunning{Size and albedo of comet C/\bb (Bernardinelli-Bernstein)  }
   \authorrunning{Lellouch et al.}

   \date{Received...; revised...}
 
  \abstract
   {The recently announced Oort-cloud comet C/2014 UN$_{271}$ (Bernardinelli-Bernstein) is remarkable in at least three respects: (i) it was discovered inbound as far as $\sim$29 au
from the Sun (with prediscovery images up to $\sim$34 au); (ii)
it showed cometary activity already at almost 24 au; and (iii) its nuclear magnitude ($H_r$ $\sim$ 8.0) indicates an exceptionally large object. Detection of
gases is expected in the upcoming years as the comet is heading towards a $\sim$11 au perihelion in 2031.}
   {The goal is to determine the object's diameter and albedo from thermal measurements.}
   {We used ALMA in extended configuration (resolution $\sim$0.064") to measure the 1287 $\mu$m (233 GHz) continuum flux of the comet. Observations were performed on August 8, 2021
at a 20.0 au distance from the Sun. The high spatial resolution
was chosen in order to filter out any dust contribution. We also used a recently published $Af\rho$ value to estimate the dust production rate and the expected dust thermal
signal for various assumptions on particle size distribution.}
   {We detected the thermal emission of the object at $\sim$10 $\sigma$, with a flux of 0.128$\pm$0.012 mJy. Based on observational constraints and our theoretical estimates of the dust contribution, the entirety of the measured flux can be attributed to the nucleus. From NEATM modelling combined with the $H_r$ magnitude, we determine a surface-equivalent
diameter of 137$\pm$17 km and a red geometric albedo of 5.3$\pm$1.2~\%. This confirms that C/\bb\ is by far the largest Oort-cloud object ever found (almost twice as large as comet C/1995 O1 Hale-Bopp), and except for the Centaur 95P/Chiron which shows outburst-like activity, the largest known comet in the Solar System. On the other hand, the C/\bb\ albedo is typical
of comets, adding credence for a ``universal" comet nucleus albedo.}
   {With its distant perihelion and uniquely large size, C/2014 UN$_{271}$ (Bernardinelli-Bernstein) is the prominent archetype of distant comets, whose activity is driven by hypervolatiles. Monitoring of dust and gas emission as the comet will approach and pass perihelion will permit to study its activity time pattern and compare it to the distant
(outbound) activity of Hale-Bopp. Post-perihelion thermal measurements will permit to study possible albedo changes, such as a surface brightening compared
to pre-perihelion, as was observed for Hale-Bopp.}


   \maketitle
%
%-------------------------------------------------------------------

\begin{table*}
\caption{Observation parameters} 
\label{observations}     
\centering                          
\begin{center}
\begin{tabular}{|c|c|c|cc|cc|c|c|}
\hline
Scheduling & UT Date   & Integration   & \multicolumn{2}{|c|}{Flux calibrator}  & \multicolumn{2}{|c|}{Phase calibrator} & \multicolumn{2}{|c|}{\bb} \\
block &  (start/end) & time  & \multicolumn{2}{|c|}{and flux density$^a$} & \multicolumn{2}{|c|}{and flux density$^b$}  & Flux density$^b$ & ($\Delta$RA, $\Delta$DEC)$^c$ \\

\hline
& & & & & & & & \\
SB1 & 08-Aug-2021  & 1624 sec  & J2258-2758 & 1.22 Jy & J0253-5411 & 0.374 Jy & 0.109$\pm$0.016 mJy & -0.091" ~ -0.249"  \\
& 7:37 - 8:35 & & & & & & & \\
SB2 & 08-Aug-2021  & 1613 sec   & J0519-4546  & 1.26 Jy  & J0253-5411  & 0.389 Jy & 0.147$\pm$0.016 mJy & -0.100" ~ -0.262" \\
& 8:35 - 9:30 & & & & & & & \\
\hline
\multicolumn{9}{l}{\footnotesize $^a$ Assumed; $^b$ Measured;} \\
\multicolumn{9}{l}{\footnotesize $^c$ (RA, DEC) offset with respect to predicted position. Adopting a mean (-0.0955", -0.2555") offset for the two SBs, the J2000 astrometric coordinates   } \\
\multicolumn{9}{l}{\footnotesize with respect to the observing site (Code -7) on August 8, 2021, UT = 8:00 are: RA = 02:42:04:2508 and DEC = -53:26:48.520. } 

\end{tabular}
\end{center}
\end{table*}

\section{Introduction}
Comet C/\bb\ (Bernardinelli-Bernstein) (hereafter \bb\ for brevity) was discovered as part of the search for outer solar system objects with the Dark Energy Survey (DES) performed over 2013–2019 \citep{bernardinelli21a}. \bb\ was observed in 42 DES survey images on 25 nights over Oct. 2014 - Nov 2018, with a heliocentric distance r$_h$ $\sim$ 29 au \citep{bernardinelli21b}.
Prediscovery images from WISE, CFHT, VST, VISTA and PanSTARRS extend the photometric record to Oct. 2010 (r$_h$~$\sim$~34.1 au). Orbital analysis \citep{bernardinelli21b} indicates
characteristic Oort-cloud membership, with inclination and semi-major axis of the incoming orbit of  95.5\dg\ and 20,200 au respectively (i.e., an inbound orbital
period of $\sim$2.9 million years). The object is heading towards a 10.95 au perihelion passage on 21 Jan. 2031. Backward orbit integration points to a previous perihelic passage at 17-21 au, and suggests that \bb\ has never been closer than this distance since its ejection from the Oort cloud, possibly making it one of the most `pristine' comets ever observed.

The announcement of the object on 19 June 2021 \citep{bernardinelli21c} prompted immediate observations, that showed a visible coma at $r_h$ = 20.18 au \citep{demetz21,kokotanekova21,buzzi21}. Analysis of TESS data from Sept.-Oct. 2018 indicated that a coma was already present at 23.8 au, and syndyne analysis suggested
the activity might have started several years earlier \citep{farnham21}.
\bb\ thus joins the list of inbound, distantly active, long-period comets, that includes C/2017 K2 (PanSTARRS), C/2010 U3 (Boattini), and C/2014 B1 (Schwartz). Also, comet C/1995 O1 Hale-Bopp was observed to be active outbound 11 years after perihelion at 25.7 au from the Sun \citep{szabo08}, and may be even at 30.7 au \citep{szabo11}. Activity in these objects may in fact occur even farther:  modelling of the r$_h$ dependence of the dust production rate in C/2017 K2 indicates activity was already present at 35 au, presumably driven by the sublimation of CO or other supervolatile ices \citep{jewitt21}.

Of further interest for the characterization of \bb\ are the following aspects \citep{ridden21,farnham21, bernardinelli21b,kokotanekova21}: (i) possible fluctuations of
the magnitude, but a lack of a clear rotational signal at the $>$ 0.2-0.3 mag level; (ii) a moderately red color, with spectral slope 5-10 \% / 100 nm, typical of (or slightly bluer than) long-period comets \citep{jewitt15}. Gas emissions have not been detected yet, with production rates Q$_{CO}$ $<$ 1.25$\times$10$^{28}$ mol s$^{-1}$ and Q$_{CN}$ $<$ 1.25$\times$10$^{27}$ mol s$^{-1}$ near $\sim$21 and $\sim$20 au respectively \citep{farnham21,kokotanekova21}. Most remarkable for \bb\ is however
its exceptionnally bright absolute magnitude $H_r$ = 7.96$\pm$0.03, which showed no evolution in ground-based data during the approach from 34 to 22 au and therefore presumably
represents the nuclear magnitude \citep{bernardinelli21b}. 
For a standard cometary  5~\% red albedo, and based on usual relationships between diameter $D$ and $H$ magnitude (see Sect.~\ref{model}),  this
yields $D$ = 130 km, potentially 1.75 times larger than Hale-Bopp \citep[74$\pm$6 km diameter;][]{szabo12}. \bb\ would thus compete in size
with the largest Centaurs and with the smallest of the TNOs having size measurements, whose typical (visible) albedos are however more like 8 \% \citep{muller20}. Based on Hale-Bopp experience \citep{biver02,rauer03}, detection of CO, CN and other species in \bb\ may be expected on its way towards and beyond its $\sim$ 11 au perihelion in the upcoming years, and a determination of its diameter and albedo is needed to complete its physical characterization. We report here on such measurements based on the detection of thermal emission with 
ALMA and the application of the radiometric technique.

   \begin{figure}
   \includegraphics[angle=0,angle=90,width=9cm]{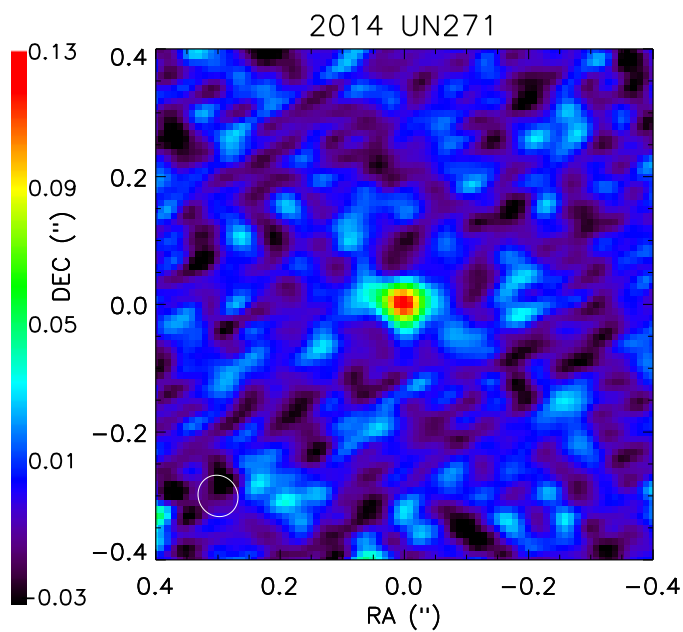}
   \caption{Recentered 233 GHz total image of \bb\, merging data from the two SB and the four spectral windows. The synthesized beam, shown in white, is 0.067''$\times$ 0.062''. The scale of the color bar is in mJy.}
   \label{fig:alma_images}
    \end{figure}

%--------------------------------------------------------------------
\section{Observations and data reduction}\label{sec:obs}
Observations of comet C/2014 UN$_{271}$ were obtained on August 8, 2021, with the 12-m array of the Atacama Large Millimeter Array (ALMA), under the ALMA DDT project 2019.A.00038
(see details in Table~\ref{observations}). 
Heliocentric distance, geocentric distance, and phase angle values of the target were $r_h$ = 20.0046 au, $\Delta$ = 19.6776 au and $\alpha$ = 2.77\dg, respectively.
All observations were taken in the ALMA Band 6 (211-275 GHz), in the continuum (``TDM") mode. We used the standard frequency tuning
for that band, yielding four 1.875-GHz broad spectral windows centered at 224, 226, 240 and 242 GHz.
The array was in extended configuration C8, with 40 operating antennas. This yielded baselines in the range 92m -- 8282m, and a synthesized beam of $\sim$0.065"  for robust
weighting (0.5), 
larger than the expected nucleus size ($\sim$10 mas) but allowing to filter out dust emission.
Observations were obtained in dual polarization mode, with the two polarizations combined at data reduction stage to provide a measurement of the total flux.

The observations consisted of two scheduling blocks (SB), each $\sim$55-58 min long including $\sim$27 min on source. The rest
of the time in each scheduling block was spent on flux/bandpass calibrators, and on secondary (phase) calibrators 
for monitoring the atmospheric and instrumental amplitude and phase gains. Observations occured in excellent weather conditions (zenithal precipitable water vapor $\sim$ 0.9 mm; antenna-based phase rms $\sim$ 26 degrees). Observational details and flux results are given in Table~\ref{observations}.

The flux/bandpass calibrators, namely quasars J0519-4546 and J2258-2758 for the two SBs, are actually variable, but routinely monitored. 
Details on how we estimated their flux and associated error bar on the observing date 
are given in Appendix~\ref{sec:calibration}. 

Initial steps of the data reduction were performed in the CASA reduction package via the ALMA pipeline \citep{muders14}, 
providing a set of visibilities as a function of baseline between each antenna pair. Visibility fitting was performed for the two flux calibrators, and visibilities
were rescaled (by factors 0.99 and 0.995 for SB1 and SB2, respectively, from the ALMA pipeline calibration) so that their measured flux matched the values expected from Appendix~\ref{sec:calibration} (and recalled in Table~\ref{observations}).
These factors were also applied to visibility data of the phase calibrator (J0253-5441) and of the science target. Visibility fitting provided two independent values
of the flux of J0253-5441, found to be consistent within 4 \% (see Table~\ref{observations}), confirming the quality of the flux scale.

For \bb, visibilities were exported into the GILDAS package for imaging and visibility fitting.
Combining data from the two SBs and the four spectral windows, a 233-GHz image of C/2014 UN$_{271}$ is shown in Fig.~\ref{fig:alma_images}, yielding
a detection of the object at $\sim$ 10~$\sigma$. 
Visibility fitting was performed independently for the two SB. In a first step, for each SB, each of the four spectral windows yielded a measure of the flux and of the (RA, DEC) position offset of the target from the expected ephemeris\footnote{JPL/Horizons interrogated on July 14, 2021}, letting these parameters free in the fit and specifying a point-like source. The four sets 
of (RA, DEC) values were then kept fixed at their average values (see Table~\ref{observations}), and the visibilities were refit in terms of the object total flux. We specified either a point-like source or
a 9.5 milli-arcsec (mas) disk (which corresponds to the a posteriori determination of the object diameter), with insignificant differences ($\sim$ 0.001 mJy in flux). 
For each SB, the combined 233 GHz flux and its error bar were obtained by merging the visibilities from the four spectra windows (GILDAS task \verb uv_merge ), after 
rescaling to that frequency using a spectral index $\alpha$ = d(log $F_\nu$)/d(log $\nu$) of 1.93 in the 224-242 GHz range, as expected from NEATM models (see Sect.~\ref{model}), and performing again visibility fitting. This yielded 0.109$\pm$0.016 mJy for SB1 and 0.147$\pm$0.016 mJy for SB2, i.e. a somewhat unexpected 2.4-$\sigma$ difference. In theory, the object's thermal flux might show rotational variability associated with a triaxial ($a$, $b$, $c$) projected shape,  
but even assuming the two SB were precisely in phase with projected surface maxima and minima, the 35 \% higher flux in SB2 would strictly imply $a$ $>$ 1.8 $b$. This is also at face value inconsistent with the lack of rotational variability at the $>$0.2--0.3 magnitude level in optical data, although the latter might be damped by coma contribution. In what follows, we simply averaged the fluxes from the two SB, yielding 0.128$\pm$0.011 mJy~\footnote{This error bar is also satisfactorily consistent with $stdev$/$\sqrt{8}$ = 0.013 mJy where $stdev$ is the standard deviation between the 8 individual flux values rescaled to 233 GHz.}. Quadratically adding a conservative 4 \% uncertainty on the flux calibrator scale, the final object flux is 0.128$\pm$0.012 mJy.  \\

\section{Analysis}
\subsection{Estimate of coma contribution}
\begin{figure}
%   \centering
   \includegraphics[angle=90,width=9cm]{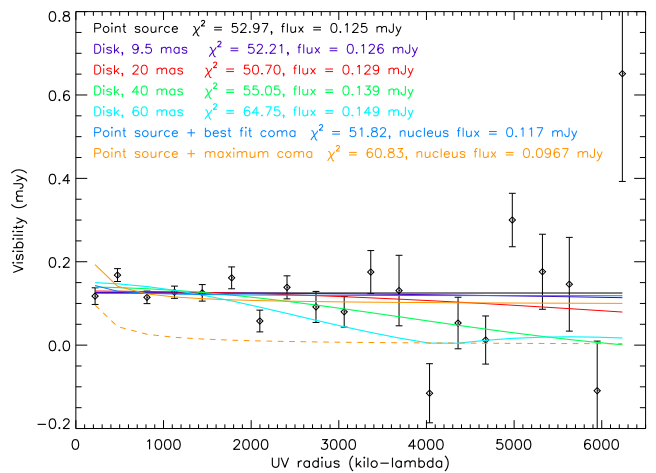}
   \caption{Real part of the visibilities, averaged in 400-m UV radius bins, compared to various models: (i) point source (ii) disks with 9.5, 20, 40 mas and 60 mas apparent size, and (iii) sum of point source and a coma. The visibilities are rescaled to 233 GHz and the UV radii are expressed in terms of the wavelength (kilo-$\lambda$). The orange dashed curve indicates the strongest coma signal that can be accomodated by the data.     }%
   \label{fig:fit_uv}
    \end{figure}

Visibility curves $V$ as a function of UV radius $\sigma$ bear information on the spatial distribution of the source.
For a uniform disk of apparent diameter $\theta$, V($\sigma$) follows a J$_1$ Bessel function, with first zero at $\sigma_0$ = 1.22 $\lambda$/ $\theta$,
and a constant value for a point-like source. Extended emission for a coma brightness distribution varying as 1/$\rho$, where $\rho$ is the distance to comet center,
shows up as visibilities V($\sigma$) $\propto$ 1 / $\sigma$ \citep{bockelee10}.
Fig.~\ref{fig:fit_uv} shows the real part of the observed visibilities for \bb, weight-averaged in 400-m wide $\sigma$ bins, the latter being plotted in terms of $\lambda$. 
From $\chi^2$ analysis, visibility fitting cannot distinguish between disk sizes smaller than $\sim$40 mas, which is to be expected given the $\sim$65 mas resolution
achieved and the measurements' S/N. Most importantly, within noise level, no indication of a 1 / $\sigma$ signal component is apparent. Fitting the V($\sigma$) curve with the sum of a constant and a 1 / $\sigma$ term provides no better fit than the previous point-like and $<$ 40 mas disk models, and in this case (sixth model in Fig.~\ref{fig:fit_uv}, blue curve), the nucleus contributes 93 \% of the signal. A 3$\sigma$ upper limit to the contribution of the coma to the visibility at $\sigma$ = 270 m (220 k$\lambda$) can be set at the 0.1 mJy level (orange curves in Fig.~\ref{fig:fit_uv}). Even in this extreme case, the contribution of the nucleus is 0.097 mJy, i.e. 76 \% of the total flux. This is a first, observational, proof of a dominant nuclear contribution to the signal. 

Furthermore, based on the comet dust activity level reported on 29 June 2021 \citep{Dekelver21,bernardinelli21b}, i.e. an $Af\rho$ value of $\sim$150 m, we estimated the expected thermal signal from the dust in the ALMA synthesized beam, for different assumptions of the particle size index ($\beta$) and maximum particle radius  ($a_{\rm max}$). Results, described in Appendix~\ref{signal:dust}, indicate that for most assumptions, the thermal emission of dust is entirely insignificant. Only if the particle size distribution is extremely shallow  ($\beta$ = 3) and the maximum particle size very large ($a_{\rm max}$ = 1 cm) is there a non-negligible dust contribution to the measured signal, but still minor
and within the uncertainty of the measured visibilities.  In the rest of the paper, the measured thermal flux of 0.128$\pm$0.012 mJy is attributed to the nucleus only.

\subsection{Nucleus diameter and albedo}
\label{model}
In the lack of knowledge of nucleus shape and spin parameters (pole orientation and shape), a thermophysical model is pointless, and we instead adopted a NEATM model, used extensively for asteroids \citep{harris98} and trans-neptunian objects \citep[][and references therein]{muller20}. NEATM is based on the asteroid standard thermal model \citep[STM;][]{lebofsky89}, but accounts for phase angle effects; additionally, the temperature distribution is modified by an adjustable $\eta^{-1/4}$ factor, representing the combined and opposed effects of roughness ($\eta$ $<$ 1) and thermal inertia ($\eta$ $>$ 1). For fixed surface (thermal inertia, roughness) and spin properties, $\eta$ is also a function of the subsolar temperature,
i.e. the heliocentric distance \citep[e.g.][]{spencer89,lellouch13}.
Given the $r_h$ = 20 au distance of our measurements (and the expected large size of \bb), we adopted a beaming factor $\eta$ = 1.175$\pm$0.42, based on measurements of 85 Centaurs and TNOs \citep{lellouch13,lellouch17}. We also specified a bolometric emissivity $\epsilon_b$  = 0.90$\pm$0.06 and a {\em relative} radio emissivity $\epsilon_r$ = $\epsilon_{mm}$ / $\epsilon_b$ = 0.70$\pm$0.13, as inferred from combined Spitzer/Herschel/ALMA measurements of 9 objects \citep{brown17,lellouch17}. The lower-than-unity relative radio emissivity is interpreted
as resulting from (i) the sounding of a colder dayside subsurface and (ii) the loss of outgoing thermal radiation due to volume scattering in the subsurface and/or
Fresnel reflection at the surface.  
The few available radio-observations of cometary nuclei also generally indicate radio-emissivities lower than 1, e.g $\sim$0.5 for Hale-Bopp \citep{fernandez02}, and $<$0.8 for 8P/Tuttle \citep{boissier11}. Comets are also found to have low thermal inertias \citep[e.g., $<$10, $<$30 and $<$45 MKS for 8P/Tuttle, 22P/Kopff, and 9P/Tempel 1, respectively;][]{boissier11,groussin09,groussin13}, consistent with a beaming factor $\eta$ of order unity. Based on NEATM analysis of a large sample of comet nuclei observed
with Spitzer at $r_h$ = 3.5--6 au, \cite{fernandez13} find a mean $\eta$ of 1.03$\pm$0.11; and the large (D = 65 km) 29P/Schwassmann-Wachmann nucleus has $\eta$ = 1.1$\pm$0.2
\citep{2021PSJ.....2..126S}. These numbers are fully consistent with our above choice of $\eta$. Given the values of r$_h$, $\eta$, and $\epsilon_b$, NEATM calculations indicate that the object's spectral index over 224-242 GHz is 1.93, sligthly lower than the Rayleigh-Jeans limit of 2.

With the above parameters, the measured thermal flux yields the object's (surface-equivalent) diameter $D$, and the albedo is then determined from the usual relationship between diameter $D$ and $H$ magnitude, i.e. $D$ = 2 $a$ / $\sqrt{p}$ 10$^{0.2 (m_{\odot} - H)}$, where $p$ is the object's geometric albedo and $m_{\odot}$ the solar magnitude in the relevant band, and $a$ = 1 au. Using V$_{\odot}$ = -26.76 and (V--R)$_{\odot}$ = 0.35, one obtains $D$ = 1330 km /$\sqrt{p_V}$ 10$^{-0.2 H_V}$ from V band and $D$ = 1132 km /$\sqrt{p_R}$ 10$^{-0.2 H_R}$ from R band. We converted the {\it griz} magnitudes and colors from \citet{bernardinelli21b} using the prescriptions of \citet{jester05}, yielding 
$H_V$ = 8.21$\pm$0.05, $H_R$~=~7.75$\pm$0.05, and V-R = 0.46$\pm$0.02, i.e. $p_R$ = 1.11$\pm$0.02 $p_V$.
The temperature distribution $T$ across the object is also a function of the object's albedo $p_V$ through $T$ $\propto$ (1 - $p_V~q$)$^{1/4}$ (where a reasonable value of $q$, the phase integral, is $\sim$~0.4), but this dependence is minimal given the a posteriori low albedo inferred ($p_V$ $\sim$ 0.049), so iterating once on the albedo in NEATM 
was sufficient.

To account for uncertainties, both on the measured object's flux and on the model inputs ($\eta$, $\epsilon_b$, $\epsilon_r$ and H$_r$), we randomly generated a large set (40,000) of
synthetic data and model inputs, based on gaussian-added noise on each parameter at the appropriate level \citep{mueller11}, solving for $D$ and $p_R$
in each case.  In doing so, we restricted $\eta$ values to be $>$0.6  \citep[lower values are physically implausible, see][]{mommert12}, as well as $\epsilon_b$ $<$ 1 and $\epsilon_r$ $<$ 1.

\label{analysis}

   \begin{figure}
%   \centering
   \includegraphics[angle=90,width=9cm]{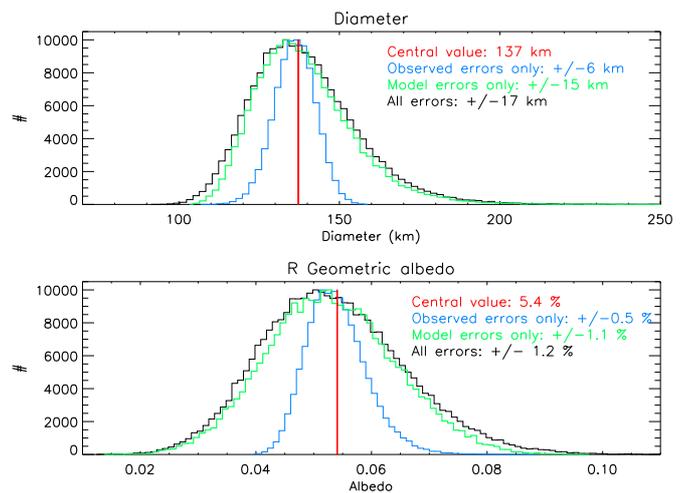}
   \caption{Diameter and R albedo distribution of solutions. Red: central values. Blue, green, and black curves show distributions associated with flux uncertainties, model uncertainties, and both. For easier comparison, all distributions are rescaled to peak at 10,000. }%
   \label{fig:solutions}
    \end{figure}

\section{Results and discussion}
Fig.~\ref{fig:solutions} shows the distribution of the solution $D$ and $p_R$, accounting separately for flux uncertainties, model uncertainties, and both. Those
provided best fit values and error bars (defined to include the central 68.3 \% of the results) for the diameter and albedo. We find $D$ = 137$\pm$17 km \footnote{This yields an apparent diameter of 9.6 mas at $\Delta$ = 19.68 au, consistent with the point-like appearance of the object in the visibility curve (Fig.~\ref{fig:fit_uv}).} and $p_R$ = 5.3$\pm$1.2 \%, where error bars are dominated by model errors -- and principally by the uncertainty on $\epsilon_r$. This confirms that \bb\ is almost twice larger in diameter than 
Hale-Bopp and makes it the largest Oort Cloud comet detected so far. \bb\ is also larger than almost all active Centaurs \citep[e.g. D$\sim$60 km for 167P/Cineos, 174P/Echeclus, and 29P/Schwassmann-Wachmann, see][]{muller20,2021PSJ.....2..126S}, being surpassed only by 95P/Chiron \citep[D$\sim$215 km;][]{fornasier13}. Given however that
Centaurs activity occurs mostly in the form of outbursts of variable lifetime\footnote{In the case of 29P, superimposed on a steady background activity level.} but is mostly uncorrelated with heliocentric distance \citep[][and references therein]{peixinho20}, \bb\ appears as the largest ``standard comet" ever discovered\footnote{For \bb, \cite{kelley21} reported activity variation possibly associated with outbursts, but the long-term behavior is more characteristic of continuous activity \citep{farnham21}.}.

Unlike the size, the albedo of \bb, $p_V$ = 4.9$\pm$1.1 \%, is completely in line with that of other, typically much smaller, comets \citep[2--6 \% in V or R for a sample of 
$\sim$80 ecliptic or near-isotropic comets, with no discernible trends with other orbital or physical parameters, see][]{lamy04, campins02}. Our measurement thus adds evidence against a dependence of comet nucleus albedo on size. Such a conclusion was reached previously by \citet{fernandez13} on the basis that the size distribution they measured for 89 comets in the thermal range is indistinguishable from that inferred from optical photometry assuming constant albedo. In this context, 95P/Chiron, with $p_V$ = 0.10-0.17 \citep{lellouch17}, stands as a clear outlier both from comet nuclei and from the Centaur population as a whole \citep[median $p_V$, 5.6 \%; see][]{muller20} \footnote{We note that the geometric
albedo of 29P is very uncertain, $p_V$ =~2.5-12~\%, depending on the adopted H$_v$ magnitude \citep{stansberry04}.}.
We also note that with its $\sim$5 \% albedo and 5-10 \% /100 nm spectral slope, \bb\ falls in the middle of the ``dark/neutral" cluster identified in the Kuiper Belt \citep{lacerda14}. The low albedos encountered on many outer solar system objects are usually associated with the presence of exposed organics, along with additional darkening agents like sulfides \citep{rousseau18}. 
This hypothesis is strengthened by the detection of large amounts ($\sim$50 \% in mass) of organics in the dust of 67P/Churyumov-Gerasimenko \citep{bardyn17}, but many questions remain as to the relations between albedo, color, composition, irradiation, and activity \citep[see e.g.,][]{brunetto06,jewitt15,poston18,wong19}.

Our observation at 20 au provides the most distant determination of the albedo of a new Oort Cloud object on its inbound orbit. This is of interest because cometary activity may cause nucleus albedo (and color) to change over time.  
In the case of comet Hale-Bopp, a joint analysis of pre- and post-perihelion data indicated $p_R$ $\sim$ (3.1-3.6)$\pm$1.0  \% at 6.4 and 4.4 au inbound \citep{szabo12}, in
agreement within errors with our determination for \bb, but an exceptionally high $p_R$ = 8.1$\pm$ 0.9 \% at 31-32 au outbound. This was intepreted as due to gravitational redeposition of bright icy grains near the cessation of the outbound activity, burying low-albedo material. Although the mechanism was at the time deemed more likely to occur in large objects (favoring gravitational fall-back) and with distant activity (associated with slower velocities), a similar redeposition mechanism was responsible for the
bright, smooth, ejecta-covered ``neck'' (Hapi) region of comet 67P/Churyumov-Gerasimenko\footnote{The nucleus of 67P also underwent blueing and brightening during the perihelion passage, due to the blowing-off of volatile-depleted superficial layers and the exposure of brighter and bluer ice-rich sub-surface layers \citep{fornasier16,filacchione20}.}.
Remeasuring the thermal emission (and colors) of \bb\ post-perihelion (e.g. at 20 au outbound in 2040) will permit to assess whether these processes occur as well on this comet.

Just as Hale-Bopp is the archetype of a large comet on a Sun-approaching orbit, \bb\ appears as the most prominent representative of distant, long-period comets, whose activity is governed by hypervolatiles (CO, CO$_2$,...), and monitoring of its chemical composition as it will approach and pass perihelion will be of high value.
Scaling Hale-Bopp (outbound) activity data \citep{biver02} by D$^2$ and r$_h^{-2}$, we expect a current CO production rate in \bb\ of Q$_{CO}$  = 7$\times$10$^{27}$ mol s$^{-1}$, rising to $\sim$2$\times$10$^{28}$ mol s$^{-1}$ at the $\sim$11 au perihelion in January 2031. Likewise, based on Hale-Bopp CN data up to 9.8 au \citep{rauer03},
we anticipate Q$_{CN}$ $\sim$ 2$\times$10$^{25}$ mol s$^{-1}$ at perihelion. While signals will remain modest, requiring the use of sensitive facilities
(ALMA, VLT, JWST...), both species and possibly a few others (HCN, CH$_3$OH, CO$_2$...), should be detectable and monitored over a $\sim$10 year period around perihelion.
The biggest difference between \bb\ and Hale-Bopp, however, is that the former will not enter the water-dominated activity regime, and comparisons between  
the instrinsic (i.e., per km$^2$) activity pattern (outbound, for Hale-Bopp) in the two comets, and possibly in some active Centaurs, will provide further insights into the mechanisms of distant cometary activity. Additional information on the spin properties, shape, and thermal regime of \bb\ should also be gained in the near future, from combined optical
imaging, JWST thermal measurements, and possibly stellar occultations.

\begin{acknowledgements}
This paper is based on ALMA program 2019.A.00038. ALMA
is a partnership of ESO (representing its member states), NSF
(USA) and NINS (Japan), together with NRC (Canada), NSC and
ASIAA (Taiwan), and KASI (Republic of Korea), in cooperation with
the Republic of Chile. The Joint ALMA Observatory is operated
by ESO, AUI/NRAO and NAOJ. The National Radio Astronomy
Observatory is a facility of the National Science Foundation operated under cooperative agreement by Associated Universities,
Inc. P.S-S. acknowledges financial support from the Spanish grant AYA-RTI2018-098657-J-I00 ``LEO-SBNAF'' (MCIU/AEI/FEDER, UE) and
from the State Agency for Research of the Spanish MCIU through the ``Center of Excellence Severo Ochoa'' award to the
Instituto de Astrofísica de Andalucía (SEV-2017-0709).
\end{acknowledgements}


\begin{appendix}
\section{Absolute flux calibration}
 \label{sec:calibration}
\begin{figure}[ht]
 \label{fig:calibration}
\centering
\includegraphics[width=\columnwidth,angle=0]{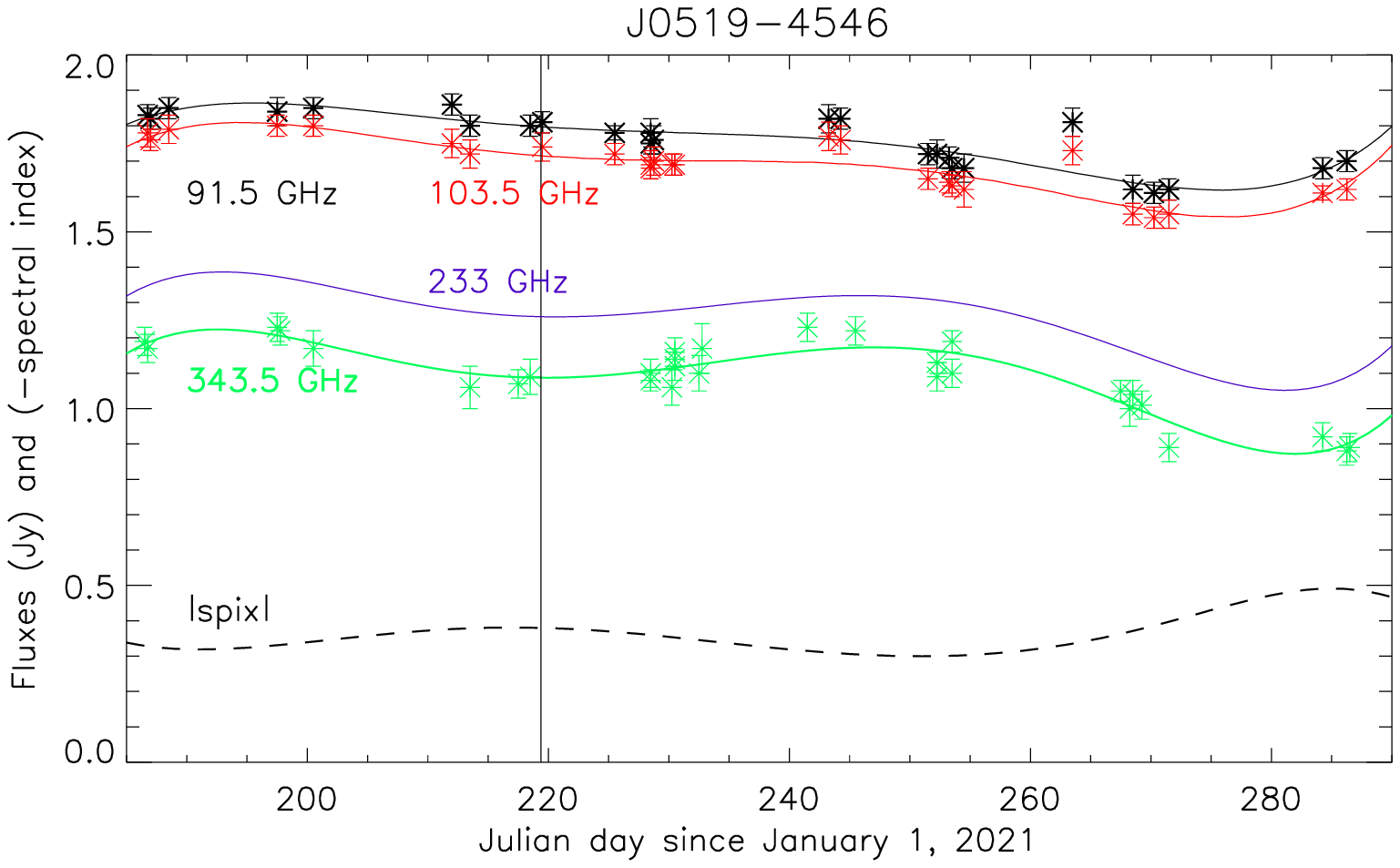}
\includegraphics[width=\columnwidth,angle=0]{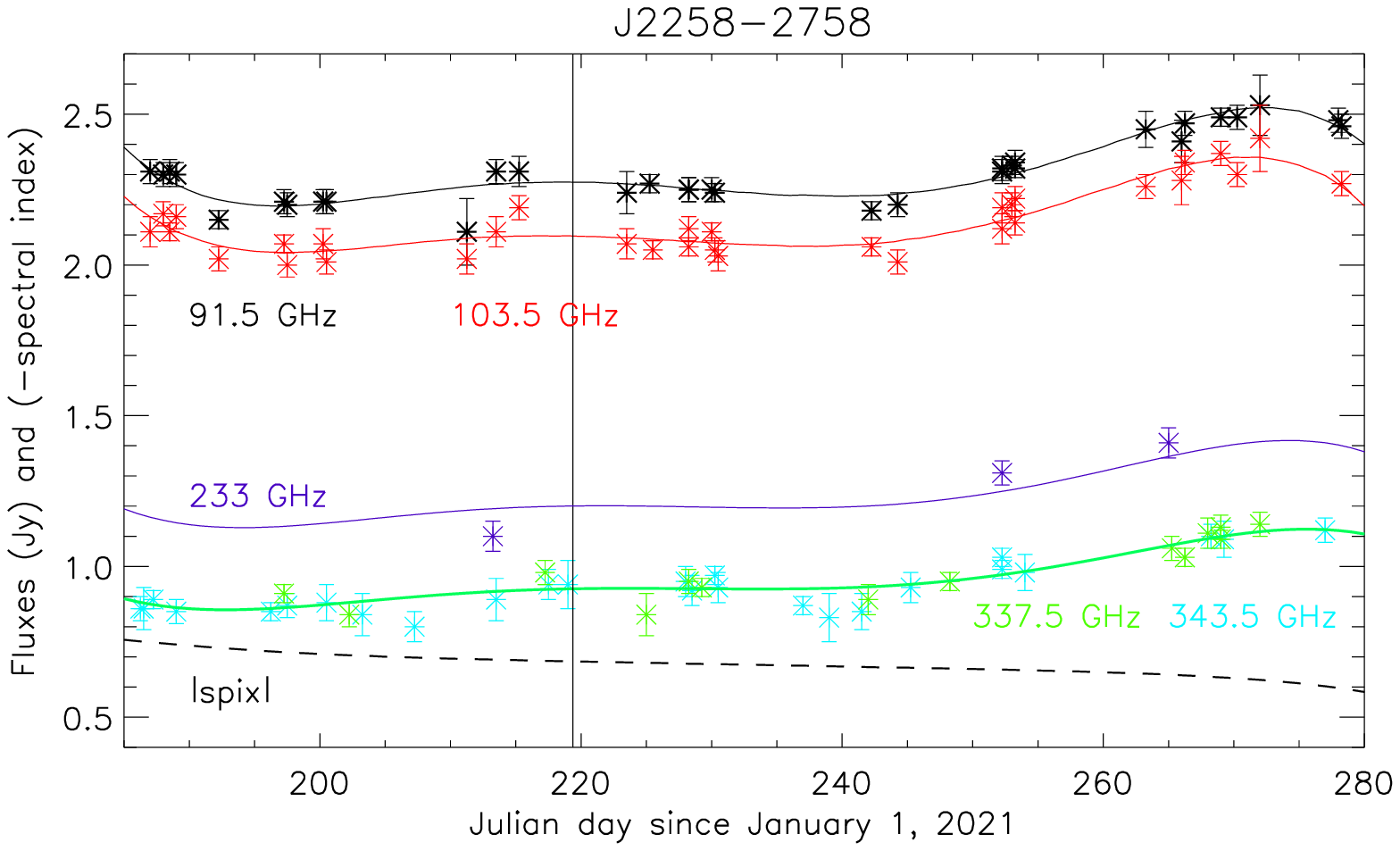}
\caption{Flux measurements of the two flux/bandpass calibrators, over a $\sim$100 day period  spanning our observation period. These measurements, color-coded per frequency, are taken from the ALMA Calibrator Source Catalogue at 91.5, 103.5, and 337.5 or 343.5 GHz. The solid lines are polynomial fits, from which spectral indices
and the 233 GHz flux as a function of time are inferred. In the case of J2258-2758, a few direct 233 GHz measurements are also available and plotted. The dashed line, labelled "|spix|", shows the absolute value of the spectral index. The vertical bar corresponds to the date of the \bb\ observations.
}
\end{figure}

A good knowledge of the calibrators is critical for the reliability of the flux scale. Our flux/bandpass calibrators, J2258-2758 and J0519-4546 for the first and second scheduling blocks (SBs) respectively, are actually variable, but routinely monitored from ALMA\footnote {https://almascience.eso.org/sc/}, 
mostly at 91.5, 103.5, and $\sim$ 340 GHz (337.5 or 343.5 GHz), and for the second one occasionnally at 233 GHz. 
These measurements are displayed in Fig.~\ref{fig:calibration}.1 over a $\sim$100 day period spanning our observing period. We fitted the 91.5, 103.5, and $\sim$ 340 GHz fluxes with 5-degree polynomials, from which the spectral index over this frequency range and the 233 GHz flux was determined as a function of time. Interpolation on the date of our observation then yielded
the desired quasar 233 GHz flux. This approach, illustrated in Fig.~\ref{fig:calibration}.1, yielded 1.26 Jy for J0519-4546 and 1.20 Jy for J2258-2758 on August 8, 2021.
For the latter object, the value is also nicely consistent with the few available 233 GHz measurements. Moreover, J2258-2758 is also monitored at $\sim$225.5 GHz from the SMA
\footnote{http://sma1.sma.hawaii.edu/callist/callist.html}. Interpolating between 225.5 GHz measurements from July 6, July 9 and August 17, 2021, and applying a spectral index of -0.68
(see Fig. \ref{fig:calibration}.1) leads to a 233 GHz flux of 1.24 Jy, comparing well with the above 1.20 Jy value. We finally adopt a 1.22 Jy flux for J2258-2758, with an estimated 2 \% uncertainty.

As indicated in the main text, each of the two scheduling blocks used the same phase calibrator, J0253-5441. The latter is not monitored in terms of flux at ALMA.
However, calibrating the visibilities on each flux/bandpass calibrator yielded two independent measurements of its flux, namely 0.374 and 0.389 Jy for the two SBs
(Table~\ref{observations}). We adopt the relative difference, 4~\%, as a conservative estimate on the absolute calibration uncertainty. 
This uncertainty is dwarfed by the S/N-limited error bar on the \bb\ flux, equivalent to 9 \%, but was taken into account in the final error bar.

\section{Dust signal estimation}
\label{signal:dust}
We estimated the coma flux density at 233 GHz in the ALMA 0.067\arcsec$\times$0.062\arcsec~synthesized beam on the basis of reported optical aperture photometry. In practice we used the $Af\rho$ value of $\sim$ 150 m deduced from the observations of  C/2014 UN$_{271}$ on 29 June 2021 ($r_{\rm h}$ = 20.15 ua) \citep{Dekelver21} \citep[see also][]{bernardinelli21b}. The $Af\rho$ parameter, where $A$ and $f$ are the grain albedo and the filling factor of the instrument field of view with projected radius $\rho$, is independent of $\rho$ if the line of sight column density falls off as 1/$\rho$, as expected for steady and isotropic dust production \citep{1984AJ.....89..579A}. $Af\rho$ is proportional to the dust production rate, with a coefficient of proportionality that depends on the particle size and velocity distributions. Both the independence on $\rho$, and the increase of $Af\rho$ as \bb\ approached the Sun from 28 to 20 au, is consistent with $Af\rho$ measuring the actual activity rate of the comet \citep{bernardinelli21b}.

Our estimation of the coma flux density at 233 GHz is based on two steps: 1) the determination of the dust production rate $Q_{\rm d}$ from the measured $Af\rho$ using the Mie scattering calculations of \citet{2012Icar..221..721F}; 2) the computation of the dust thermal emission using the model of \citet{2017MNRAS.469S.443B}, also using Mie theory.  Calculations were made for size distributions given by d$Q_{\rm d}(a)$/da $\propto a^{-\beta}$, where $a$ is the particle radius and $\beta$ is the size index. We considered a minimum size of 0.01~$\mu$m, and maximum sizes $a_{\rm max}$ of 10 $\mu$m and 1 cm.

\subsection{Dust production rate}
\citet{2012Icar..221..721F} provide the $Q_{\rm d}$/$Af\rho$ ratio (kg s$^{-1}$~/~m) for particle sizes in the range 0.01~$\mu$m~--~1 cm and various size index values. Their calculations were made for a refractive index $n$ = 2.00 + 0.10$i$, a phase angle $\phi$ = 40$^{\circ}$ and particle velocities following $v$($a$) = 0.1423 $a^{-0.5}$ m/s, $a$ being the particle radius in m ($v$=142.3 m/s for $a$= 1 $\mu$m). We rescaled $Q_{\rm d}$/$Af\rho$ to a phase angle of 3$^{\circ}$, assuming a phase function ratio $p$($\phi$=3$^{\circ}$)/$p$($\phi$=40$^{\circ}$) = 2.42. This value matches the composite phase fonction of D. Schleicher\footnote{\url{https://asteroid.lowell.edu/comet/dustphase.html}} and is in the range of the values obtained from Mie calculations \citep{2012Icar..221..721F}. We also rescaled $Q_{\rm d}$/$Af\rho$ to a velocity distribution $v$($a$) = 0.083~$a^{-0.5}$ m/s. Indeed,  based on \citet{1997Icar..127..319C}, we derived $v$ = 83 m/s for $a$ = 1 $\mu$m for \bb, using a nucleus radius of 68 km, nucleus and dust densities of 500 kg m$^{-3}$ and 1000 kg m$^{-3}$, respectively, and a CO production rate of 4$\times$10$^{28}$ mol s$^{-1}$ emitted in cone of 45$^{\circ}$ half-aperture. The assumed CO production rate (1860 kg $^{-1}$) is
reasonable but somewhat arbitrary. An upper limit of 2$\times$10$^{28}$ mol s$^{-1}$ was derived from 4.6 $\mu$m NEOWISE observations at $r_{\rm h}$ = 20.92 ua \citep{farnham21}. Moreover, extrapolating the CO production rate of $\sim$ 4 $\times$10$^{27}$ mol s$^{-1}$ of comet C/1995 O1 (Hale-Bopp) measured at $r_{\rm h}$ = 14 au \citep{biver02} (assuming a $r_{\rm h}^{-2}$ dependence) and correcting for the different nucleus sizes \citep[37 km radius for Hale-Bopp,][]{szabo12}, yields an expected CO production rate at $r_{\rm h}$ = 20 au for C/2014 UN$_{271}$ is $\sim$ 7$\times$10$^{27}$ mol s$^{-1}$. Using the nominal CO production rate of 4$\times$10$^{28}$ mol s$^{-1}$, the rescaled $Q_{\rm d}$/$Af\rho$ ratio and the derived dust production rate $Q_{\rm d}$ for C/2014 UN$_{271}$ are given in Table~\ref{tab:dust} for size indexes of 3, 3.5, and 4, given the measured $Af\rho$ of 150 m
\citep{Dekelver21}.

Although we considered $a_{\rm max}$ of 10 $\mu$m and 1 cm for conservativeness, the large size of C/2014 UN$_{271}$ makes the release of large particles unlikely. With the adopted CO gas production rate of 4$\times$10$^{28}$ mol s$^{-1}$, we estimated a maximum liftable size of 8 $\mu$m following \citet{2018Icar..312..121Z}. For a CO production rate of 7$\times$10$^{27}$ mol s$^{-1}$, the maximum liftable size is only 1.3 $\mu$m. 
Calculations for $a_{\rm max}$ = 1.3 $\mu$m were not performed, but that case would obviously lead to even much smaller dust production rates and thermal flux densities 
than those reported in Table~\ref{tab:dust} for $a_{\rm max}$ = 10 $\mu$m. 

\begin{table}
\label{tab:dust} \caption{C/2014 UN$_{271}$ dust mass loss rate and flux density in ALMA beam.}
 \begin{tabular}{lcccc}
 \hline\hline\noalign{\smallskip}
 Size  & $a_{\rm max}$ & $Q_{\rm d}$/$Af\rho^a$  & $Q_{\rm d}$ (UN$_{271}$)$^b$ & $F$(233 GHz)$^c$\\
index &  & (kg/s/m) & (kg/s) & (mJy)\\
      \hline\noalign{\smallskip}
3 & 10 $\mu$m & 0.291 & 43.6 & 9.9 10$^{-6}$ \\
  & 1 cm & 56.9 & 8544 & 1.5 10$^{-2}$  \\
3.5 &  10 $\mu$m &  0.338 & 50.7 &1.0 10$^{-5}$ \\
    & 1 cm & 9.16 & 1374 & 2.8 10$^{-3}$ \\
4.0 & 10 $\mu$m & 0.425 &   63.7 & 1.1 10$^{-5}$  \\
    & 1 cm & 0.488 & 73.2 & 8.6 10$^{-5}$ \\ 
  \hline   

\multicolumn{5}{l}{\footnotesize $^a$ Rescaled from values of \citet{2012Icar..221..721F}, see text.} \\
\multicolumn{5}{l}{\footnotesize $^b$ For $Af\rho$ = 150 m \citep{Dekelver21}.} \\
\multicolumn{5}{l}{\footnotesize $^c$ In a 0.064\arcsec~beam width at $r_h$ = 20.0 ua and $\Delta$ =  19.68 ua.}  \\
\end{tabular}
\end{table}

\subsection{Thermal flux of the dust}
\begin{figure}[h]
\begin{center}
\includegraphics[width=9cm, angle = 0]{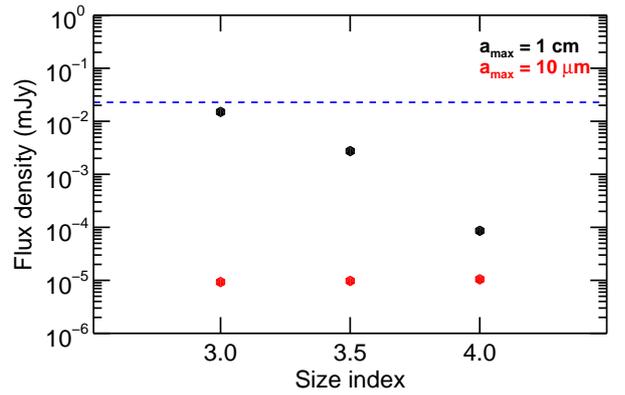}
\end{center}
 \caption{Expected flux density at 233 GHz from dust thermal emission, based on the measured Af$\rho$ of 150 m at 20.15 ua \citep{Dekelver21, bernardinelli21b}. The beam width is 0.064\arcsec. Results are presented for a size index of 3, 3.5 and 4, and for dust maximum sizes of 10 $\mu$m (red dots) and 1 cm (black dots). See text for details. The blue dashed line shows the upper limit derived from ALMA observations. }
\label{fig:dustflux}
 \end{figure}

To compute the expected dust emission at mm-wavelengths, we then used the model described in \citet{2017MNRAS.469S.443B} which computes the wavelength-dependent absorption coefficient and the temperature of the dust particles as a function of grain size, using Mie theory combined with an effective medium theory, allowing to handle mixtures of different materials. We considered a matrix of amorphous carbon with inclusions of amorphous olivine with a Fe:Mg composition of 50:50, and set the carbon/olivine mass ratio to unity. This model has also been used to analyse ALMA/ACA observations of C/2015 ER61 (PanSTARRS) and mid-infrared data of comet 29P/Schwassmann-Wachmann \citep{2021ApJ...921...14R,2021PSJ.....2..126S}. The dust local density is described as in \citet{2012Icar..221..721F}, i.e., follows a  1/($r^2$ $v$($a$)) law, where $r$ is the distance to nucleus. The flux density is computed by summing the blackbody thermal emission of individual particles over the size range, and field of view, weighted by the particle size distribution. For consistency with the model of \citet{2012Icar..221..721F}, the dust density was taken equal to 1000 kg m$^{-3}$.
The computed flux densities in a Gaussian beam of half-power beam width HPBW = 0.064" (corresponding to the ALMA synthesized beam) are given in the last column Table~\ref{tab:dust} and shown in Fig.~\ref{fig:dustflux}. We note that the derived flux values are not dependent on the adopted scaling constant in the dust velocity law, since the same velocity law
is used for the $Q_{\rm d}$/$Af\rho$ and flux calculations.

\subsection{Visibility curve for dust thermal emission}
For comparison with the measured visibility curve (Fig.~\ref{fig:fit_uv}), we also calculated the dust signal in terms of the visibility flux at 220 k$\lambda$.
In their Appendix A, \citet{bockelee10} provide formulas describing how the  amplitude of the visibility $V$ varies as a function of UV radius $\sigma$, for a brightness distribution varying as 1/$\rho$.  
This includes the formula for the visibility at $\sigma$ = 0 m, corresponding to the signal in the primary beam. The formulas are given for molecular lines, but can be applied to the dust coma, adjusting the factors describing the emission mechanism. From Eq. A.3 and A.4 of \citet{bockelee10}:
\begin{equation}
V(\sigma) = K \frac{c}{\sigma \nu}
\label{eq:first}
\end{equation}
\noindent
for $\sigma/D_{\rm beam}$ $\gg$ 0.2, where $D_{\rm beam}$ is the HPBW of the primary beam, and
\begin{equation}
V(0) = K \sqrt{\frac{\pi}{4 {\rm ln}(2)}} \pi \Phi_{beam}
\label{eq:second}
\end{equation}
\noindent
where $\Phi_{beam}$ is the HPBW of the primary beam, this time in radian\footnote{For a uniform circular aperture with same HPBW, the $\sqrt{\frac{\pi}{4 {\rm ln}(2)}}$ = 1.064 term would be replaced by 1.}.
$K$ is a constant that incorporates factors related to the emission mechanism and the distance of the comet to the observer. Equation~\ref{eq:second} is derived from Eq. A.4 of \citet{bockelee10}, here with the assumption of a Gaussian beam. Equation~\ref{eq:second} can also be used for the signal in the synthesized interferometric beam $F_{synth}$, replacing $\Phi_{beam}$ by the angular size of synthesized beam $\Phi_{synth}$. Hence,
\begin{equation}
\frac{V(\sigma)}{F_{\rm synth}} = \frac{c}{\sigma \nu }\sqrt{\frac{4 {\rm ln}(2)}{\pi}}\frac{1}{\pi \Phi_{synth}} = \frac{\lambda}{\sigma }\sqrt{\frac{4 {\rm ln}(2)}{\pi}}\frac{1}{\pi \Phi_{synth}}.
\end{equation}
From this formula, the visibility flux at 220 k$\lambda$ from the dust coma is 4.40 times the flux in a 0.064$\arcsec$ beam.

Based on Table~\ref{tab:dust} and Fig.~\ref{fig:dustflux} it appears that in virtually all cases is the expected thermal emission of dust entirely negligible. Only in one case 
(size index $\beta$ = 3, $a_{\rm max}$ = 1 cm), does the dust contribution amount to  $\sim$12 \% of the measured signal in the interferometric beam \footnote{and even less considering possible flux losses due to missing short spacing.}. Another way of seeing it is that
this extreme case corresponds to a visibility of 0.015 $\times$ 4.4 = 0.066 mJy at 220 k$\lambda$, well within the 3-$\sigma$ measured upper limit of 0.1 mJy.
The latter value corresponds to an upper limit of the dust contribution to the synthetized beam flux of 0.1 / 4.4 = 0.023 mJy, which is plotted in Fig.~\ref{fig:dustflux}.
We note finally that although \citet{farnham21} argued that the coma of \bb\ consists mostly of submillimeter sized particles emitted at low velocities, this does not appear consistent with our above estimate of the maximum liftable size. Even if this was the case, the contribution of the coma to the thermal flux would be negligible, based on
interpolation between the $a_{\rm max}$ = 1 cm and $a_{\rm max}$ = 10 $\mu$m cases in Table~\ref{tab:dust} and Fig.~\ref{fig:dustflux}.

    \end{appendix}

\end{document}